\documentclass[romanappendices,draftcls,onecolumn,12pt]{IEEEtran}
\usepackage{amsmath,amsthm}
\usepackage{graphicx}  
\usepackage{subfigure} 
\usepackage{amsmath}
\usepackage{float}
\usepackage{amssymb}
\usepackage{amstext}
\usepackage{citesort}
\usepackage[nolists,nomarkers]{endfloat}
\usepackage[endfloat,ruled]{algorithm2e}  
\usepackage{algorithmic}

\newtheorem{theorem}{Theorem}

\newcommand{\etal}{\emph{et~al.~}}

\newcommand{\bm}[1]{\boldsymbol{#1}}
\newcommand{\mat}[1]{\bm{#1}}

\newcommand{\diag}{\mathrm{diag}}

\newcommand{\vectorize}{\mathrm{vec}}
\newcommand{\sinc}{\mathrm{sinc}}

\newcommand{\openone}{\leavevmode\hbox{\small1\normalsize\kern-.33em1}}

\newcommand{\EE}{\mathbb{E}}
\newcommand{\CC}{\mathbb{C}}

\begin{document}
\title{Multipath Parameter Estimation from OFDM Signals in Mobile
  Channels} \author{Nick Letzepis\thanks{N. Letzepis and A. Grant are
    with the Institute for Telecommunications Research, University of
    South Australia, e-mail: {\tt nick.letzepis@ieee.org,
      alex.grant@unisa.edu.au}. ~~P. Alexander and D. Haley are with
    Cohda Wireless Pty. Ltd., e-mail: {\tt
      \{paul.alexander,david.haley\}@cohdawireless.com.au}.},~\IEEEmembership{Member,~IEEE},
  Alex Grant,~\IEEEmembership{Senior~Member,~IEEE},\\ Paul
  Alexander~\IEEEmembership{Member,~IEEE}, David Haley\thanks{ This
    work was supported by Cohda Wireless Pty. Ltd. and the Australian
    Research Council under grant
    LP0775036.},~\IEEEmembership{Member,~IEEE}}

\maketitle

%
%
\begin{abstract}
  
  We study multipath parameter estimation from orthogonal frequency
  division multiplex signals transmitted over doubly dispersive mobile
  radio channels. We are interested in cases where the transmission is
  long enough to suffer time selectivity, but short enough such that
  the time variation can be accurately modeled as depending only on
  per-tap linear phase variations due to Doppler effects.  We
  therefore concentrate on the estimation of the complex gain, delay
  and Doppler offset of each tap of the multipath channel impulse
  response.  We show that the frequency domain channel coefficients
  for an entire packet can be expressed as the superimposition of
  two-dimensional complex sinusoids.  The maximum likelihood estimate
  requires solution of a multidimensional non-linear least squares
  problem, which is computationally infeasible in practice. We
  therefore propose a low complexity suboptimal solution based on
  iterative successive and parallel cancellation.  First, initial
  delay/Doppler estimates are obtained via successive
  cancellation. These estimates are then refined using an iterative
  parallel cancellation procedure. We demonstrate via Monte Carlo
  simulations that the root mean squared error statistics of our
  estimator are very close to the Cramer-Rao lower bound of a single
  two-dimensional sinusoid in Gaussian noise.

\end{abstract}

%
%
\section{Introduction}

In wireless communications, reflection and diffraction of the
transmitted radio signal results in the superimposition of multiple
complex-scaled and delayed copies of the signal at the receiver.  This
type of channel is commonly referred to as a \emph{multipath
  channel}. In some instances, the multiple copies add constructively,
and in others destructively resulting in \emph{multipath fading}. When
the coherence bandwidth of the channel is smaller than the bandwidth
of the radio signal then the fading is termed \emph{frequency
  selective}~\cite{Proakis:2000}. We assume the reader is familiar
with standard wide-sense stationary uncorrelated scattering models,
for an overview see e.g.~\cite{Goldsmith:2005book}.

Orthogonal frequency division multiplexing (OFDM) is a transmission
strategy specifically designed to combat frequency selective channels
with relatively low receiver complexity
\cite{Peled:1980p964,Cimini:1985p665,Vannee:1999book}.  In OFDM, the
signal bandwidth is divided into several non-overlapping (hence
orthogonal) narrowband subcarriers where the width of each subchannel
is chosen such that it is approximately frequency non-selective. Thus
only a single tap equaliser per subchannel is required to compensate
for the multipath fading. Together with the use of the fast Fourier
transform (FFT), this results in a low complexity way to handle
frequency-selective channels. As such, OFDM is now the basis of many
current and emerging wireless communications standards,
see~\cite{Vannee:2006p445,HieDen10} for an overview. Many of these standards
are targeted for outdoor mobile applications,
e.g. 802.11p~\cite{IEEE80211p}. Mobility causes the multipath channel
(and hence frequency selectivity) to change with time. If the mobility
is fast enough compared to the symbol rate, then the channel impulse
response may vary significantly within an OFDM packet. Extensive field
trials have shown that this is indeed the case for the transmission of
802.11 OFDM signals in vehicular
environments~\cite{AleHal10}. Time-varying multipath channels such as
these are commonly termed
\emph{doubly-dispersive}~\cite{Kozek:1998p1579,Liu:2004p2583,Taubock:2007isit}.

In general, a realization of a doubly selective multipath channel
time-varying impulse response can be modeled in continuous time as
\begin{equation*}
  c(t,\tau) = \sum_{p=1}^P a_p(t) \delta(\tau-\tau_p)
\end{equation*}
where $c(t,\tau)$ is the response at delay $\tau$ to an impulse at
time $t$, where $\delta(\tau)$ denotes the Dirac-delta function. The
$a_p(t)$ are the time-varying complex amplitude (magnitude and phase)
of tap $p$, with delay $\tau_p$. The number of resolvable multipath
components is $P$. The $a_p(t)$ may aggregate many more unresolvable
multipath components, typically resulting in Ricean or Rayleigh
statistics for these parameters.

Note that for sufficiently short time durations, mobility-induced
Doppler shifts manifest as linear variations of the phase of $a_p(t)$
with time. In this paper, we consider the special case where the OFDM
packet duration is short enough such that we can model the channel as
\begin{equation} \label{eq:cir}
c(t,\tau) = \sum_{p=1}^{P} a_p e^{-j 2 \pi \nu_p t} \delta(\tau - \tau_p),
\end{equation}
where $a_p$, $\tau_p$ and $\nu_p$ respectively denote the complex
gain, delay and Doppler frequency (relative to the nominal carrier
frequency) of tap $p$. These parameters are all assumed to be
\emph{constant} over the duration of an OFDM packet. In a physical
sense, this implies that changes in the relative distance and velocity
between the transmitter, receiver and scatterers are negligible over
the duration of an OFDM packet. This model is consistent with the
geometric-stochastic model presented in~\cite{KarTuf09} for short
observation windows, and has been validated experimentally in~\cite{AleHal10}.

In this paper, we concentrate on joint estimation of $a_p$, $\tau_p$
and $\nu_p$ of the multipath components assuming perfect knowledge of
the transmitted OFDM symbols. This is a practical assumption, e.g. a
transmitted training/pilot signal, or the receiver is able to decode
the signal without error (via a forward error correction code).
Estimation of these parameters is useful in a number of areas: channel
sounding and characterisation; channel prediction; reducing channel
state information for feedback in adaptive communications; and
radar. Estimation of these parameters in the OFDM setting has been
studied previously by a number of researchers from both the
communications and radar fields.  Channel estimation via an
approximate maximum likelihood parameter search algorithm was proposed
by Thomas \etal \cite{Thomas:2003p74}. Their iterative algorithm was
based on an approximation of the maximum likelihood function, where
the multipath gain values are substituted with their least-squares
estimates. In the radar community, estimation of delay/Doppler is
vital for determination of target range and velocity. Berger
\etal~\cite{Berger:2008p2384} studied the problem of extracting the
target range/velocity information from the OFDM signal in a passive
multi-static radar system~\cite{chern:2007} using digital audio/video
broadcasted signals as illuminators of opportunity. They set up the
problem as a sparse estimation problem to use recent results from
compressed sensing~\cite{Candes:2008:SPM}. In particular, they employ
the orthogonal matching pursuit
algorithm~\cite{Needell:2008arxiv,Needell:2008asilomar}, which is in
an iterative algorithm that successively removes previously estimated
multipath components from the received signal to estimate new
components. Note that Taub\"{o}ck~\etal
\cite{Taubock:2008,Taubock:2010p255} also consider compressed sensing
to estimate the OFDM channel coefficients. However, their interest is
not in the estimation of delays/Doppler, but in the frequency/time
channel coefficients.

In this paper we begin with a continuous-time model of the transmitted
OFDM signal and derive the received matched filtered signal from first
principles. Assuming the delay-spread of the channel does not exceed
the cyclic-prefix and the pass-band of the receive/transmit filters
exceed the signal bandwidth (with negligible pass-band ripple), we
show that the resulting frequency domain channel coefficients can be
represented as the superimposition of two-dimensional (2-D) complex
sinusoids, where each 2-D frequency is proportional to the delay and
Doppler of each multipath component. Similar observations have been
made by Wong and Evans~\cite{Wong:2008:pp1601-1615,Wong:2005:p2259}
although without detailed justification. Under a similar setting they
consider estimation using only OFDM pilot symbols and propose channel
prediction algorithms based on the estimation of channel parameters
via a rotational invariance technique. Using this method, they
reformulate the problem as a one-dimensional estimation problem.

Parameter estimation of 2-D sinusoids in a general setting has been
studied extensively many years prior to the work of Wong and
Evans~\cite{Wong:2008:pp1601-1615,Wong:2005:p2259}.  Estimation
methods and the Cramer-Rao-Lower-Bound (CRLB) for the single 2-D
sinusoid case was investigated by Chien~\cite{Chien:1981thesis}.  Kay
and Nekovei~\cite{Kay:1990p1807} proposed a low complexity estimator
that operates on the phase of the noisy 2-D sample data. For the
estimation of the superposition of multiple 2-D sinusoids: Bresler and
Macovski employ a 2-D version of Prony's
method~\cite{Bresler:1986p1081}; Rao \emph{et
  al.}~\cite{Rao:1994p1795} use a similar polynomial rooting approach;
and recently Kliger and Francos~\cite{Kliger:2005p2563} consider
maximum-likelihood estimation with a maximum a-posteriori (MAP) model
order selection rule for the case where the number of sinusoids is
unknown.

In this paper we concentrate purely on the estimation of the complex
amplitude, delay and Doppler of each multipath tap, assuming the
number of taps is known. Our results can be straightforwardly extended
to the case where the number of taps is unknown using well-known
\emph{model-order selection} methods~\cite{Stoica:2004p36}. The
maximum-likelihood approach requires the solution to a
multi-dimensional nonlinear least-squares estimation
problem~\cite{Pereyra:1967p27} and hence has complexity that is
prohibitive in
practice~\cite{Bresler:1986p1081,Rao:1994p1795,Kliger:2005p2563}. We
propose a low-complexity algorithm based on a two-stage process:
first, an initial estimation; followed by a refinement proceedure.  In
the same spirit
as~\cite{Thomas:2003p74,Kliger:2005p2563,Berger:2008p2384} the initial
estimation algorithm is based on successive cancellation, whereby
multipath components are subtracted from the original signal after
they are detected. In each iteration, the delay/Doppler is estimated
using periodogram search~\cite{Chien:1981thesis,Kay:1990p1807} via a
2-D bisection algorithm. The multipath complex amplitudes are then
obtained via standard linear least-square
estimation~\cite{Boyd:2004book}. Moreover, we show that this secondary
problem can be written in terms of the \emph{ambiguity
  function}~\cite{Skolnik:2002book}. Once initial estimates have been
obtained, we then propose an iterative refinement algorithm based on
parallel cancellation. Each iteration of the refinement involves
subtracting all multipath components from the received signal except
the component of interest, which is re-estimated using the 2-D
bisection algorithm. This refinement process yields significant
improvements over the standard successive cancellation approach. We
show via Monte-Carlo simulations that this refinement algorithm
achieves performance very close to the CRLB for single 2-D sinusoid
estimation.

The remainder of our paper is organised as follows. In
Section~\ref{sec:system_model} we state the system model and derive
from first principles the received match filtered frequency domain
OFDM symbols. In Section~\ref{sec:ambiguity} we derive the transmit
signal ambiguity function. Then in Section~\ref{sec:estimator} we
present our proposed estimation algorithm and enhanced refinement
process. Simulation results are presented in
Section~\ref{sec:performance}. Finally, concluding remarks are given
in Section~\ref{sec:conclusion}.

%
%
\section{System Model} \label{sec:system_model}

Consider a $K$ subcarrier OFDM system, where packets of length $L$
OFDM symbols are transmitted. Let $\mat{X} \in \CC^{L \times K}$
denote a packet of complex OFDM symbols.  Thus $X_{l,k}$, the $l,k$th
element of $\mat{X}$, denotes the $l$th symbol transmitted on
subcarrier $k$, for $l=1,\ldots,L$ and $k = 1,\ldots,K$. In practical
OFDM systems, a certain number \emph{null} subcarriers are employed to
simplify receiver design~\cite{Vannee:1999book}. To incorporate this
feature, we let $\mathcal{K}$ denote the set of null subcarrier
indices. Thus, $X_{l,k} = 0$ for all $k \in \mathcal{K}$ and $l =
1,\ldots,L$. For all other subcarriers, i.e. $k \notin \mathcal{K}$,
we assume $X_{k,l} \in \mathcal{X}$, where $\mathcal{X} \subset \CC$
is an arbitrary complex constellation. These symbols are drawn
randomly, independently and uniformly from $\mathcal{X}$, which is
normalised to have unit average energy. Thus $\EE [|X_{l,k}|^2] = 1$
for $k \notin \mathcal{K}$ and $\EE [X_{l,k} X_{n,m}^*] = 0$ for any
$n \neq l$ or $m \neq k$. The receiver is assumed to have complete
knowledge of the transmitted symbols $X_{l,k}$, e.g. a pilot/training
signal, or from the feedback of error-free decoder decisions.

Let $x(t) = \sum_{l=1}^{L} x_l(t)$ denote the complex baseband
continuous-time transmitted OFDM signal, where
\begin{equation}
x_l(t) = \frac{1}{\sqrt{KL}}\sum_{k=1}^{K} X_{l,k} e^{j 2 \pi (k-1-\lfloor K/2 \rfloor) (t - T_{\rm cp})/T}  w(t - (l-1) T_d), \label{eq:tx_ofdm_sym}
\end{equation}
is the $l$th OFDM symbol, $T_d$ is the OFDM symbol duration (seconds),
$1/T$ is the subcarrier spacing (Hz), $T_{\rm cp} = T_d - T$
is the cyclic prefix duration (seconds), and $w(t)$ is a windowing
function such that
\begin{equation} \label{eq:window_fun}
w(t) =
\begin{cases}
\tilde{w}(t) & 0 \leq t \leq T_d \\
0 & \text{otherwise},
\end{cases}
\end{equation}
and $\int_{0}^{T_d} |w(t)|^2 \, dt = 1$. %
%
%
A simple choice of windowing function is $\tilde{w}(t) =
1/\sqrt{T_d}$. Note that the assumption $w(t) = 0$ for $t \notin
(0,T_d)$ is not necessarily required, but we assume this for
simplicity. In practice, \eqref{eq:tx_ofdm_sym} is implemented in the
discrete time domain via the inverse discrete Fourier transform
(IDFT)~\cite{Vannee:1999book}.

We assume transmit and receive filter impulse responses $g_{\rm T}(t)$
and $g_{\rm R}(t)$ respectively, and let $g(t) \triangleq
\int_{-\infty}^{ \infty} g_{\rm T}(u) g_{\rm R}(t-u) \, du $ denote
the combined transmit/receive filter response.Thus,
using~\eqref{eq:cir}, we write the overall channel response as
\begin{equation}
h(t,\tau) = \int_{-\infty}^{\infty} g(u) c(t,\tau-u) \, du  =  \sum_{p=1}^{P} a_p e^{-j 2 \pi \nu_p t} g(\tau - \tau_p). \label{eq:chan_response}
\end{equation}
Application of the overall channel response~\eqref{eq:chan_response}
to~\eqref{eq:tx_ofdm_sym} plus additive Gaussian white noise (AWGN)
yields the received continuous-time baseband signal,
\begin{equation}
y(t) = \sum_{l'} \int_{-\infty}^{\infty} x_{l'}(t-\tau) h(t,\tau) \, d \tau + z(t),
\end{equation}
where $z(t) = \int_{-\infty}^{\infty} \tilde{z}(t-u) g_{\rm R} (u) \,
du$, and $\tilde{z}(t)$ is an additive white Gaussian noise (AWGN)
process. Assuming perfect OFDM symbol synchronism, the receiver
discards the cyclic prefix and performs the matched filter to the
transmitted sinusoids, i.e.
\begin{equation}
Y_{l,k} = \frac{1}{\sqrt{KL}}\int_{T_{\rm cp} + (l-1) T_d}^{lT_d} y(t) w^*(t - (l-1)T_d) e^{-j 2 \pi (k-1-\lfloor K/2 \rfloor ) (t-T_{\rm cp})/T} \, dt, \label{eq:matched_filt_out}
\end{equation}
for $k = 1,\ldots,K$ and $l = 1,\ldots,L$. Note that in practice,
$Y_{l,k}$ is obtained via the discrete Fourier transform
(DFT)~\cite{Vannee:1999book}. We assume the pass-band of filters
$g_{\rm T}$ and $g_{\rm R}$ exceed the signal bandwidth. In addition,
we assume $\max_p \tau_p < T_{\rm cp}$ and $\max_p |\nu_p| < 1/T$, so
that inter-symbol interference (ISI) and inter-carrier interference
(ICI) can be considered negligible. Under these assumptions, in
Appendix~\ref{app:rx_matched_filt} we show that the matched filtered
output~\eqref{eq:matched_filt_out} can be written in matrix form
\begin{equation}
\mat{Y} = \mat{H} \odot \mat{X} + \mat{Z}, \label{eq:chan_model}
\end{equation}
where $\odot$ denotes the element-wise (Hadamard) product, $\mat{Y}
\in \CC^{L \times K}$ is the received matrix of filtered noisy OFDM
symbols, $\mat{Z} \in \CC^{ L \times K}$ is a matrix of independent
identically distributed (i.i.d.) zero mean complex Gaussian random
variables with variance $\sigma^2$, and $\mat{H} \in \CC^{L \times K}$
are the frequency domain channel coefficients,
\begin{equation} \label{eq:channel_coeffs}
H_{l,k} =  \sum_{p=1}^{P} a_p e^{-j 2 \pi \nu_p T_d (l-1)} e^{-j 2 \pi (k-1-\lfloor K/2 \rfloor) \tau_p/T}.
\end{equation} 
In relation to~\eqref{eq:chan_model}, we define the signal-to-noise
ratio (SNR) as $\mathsf{snr} \triangleq \EE \left[ \| \mat{X} \|^2
\right] / (L \sigma^2) = ({K-|\mathcal{K}|})/{\sigma^2}$, where $\|
\cdot \|$ denotes the Frobenius norm~\cite{Horn:1985book}.

From inspection of~\eqref{eq:channel_coeffs}, we see that it is simply
the superimposition of 2-dimensional (2-D) complex exponential
signals. We may also express $\mat{H}$ as the matrix product
\begin{equation}
\mat{H} = \mat{\Psi}(\mat{\nu}) \diag(\mat{a}) \mat{\Phi}^{\dagger}(\mat{\tau}),
\end{equation}
where $\diag(\mat{a})$ denotes a $P \times P$ diagonal matrix with diagonal
entries $\mat{a} = (a_1,\ldots,a_P)$, and
\begin{align}
\Psi_{l,p} (\mat{\nu}) &= \psi_l(\nu_p) \triangleq e^{-j 2 \pi (l-1) \nu_p T_d } \\
\Phi_{k,p} (\mat{\tau}) &= \phi_k(\tau_p) \triangleq e^{j 2 \pi (k-1-\lfloor K/2 \rfloor) \tau_p/T },
\end{align}
for $p = 1,\ldots,P$, $l=1,\ldots,L$ and $k = 1,\ldots,K$. As we shall
see later, the separation of the parameters in this matrix form will
simplify the development of our estimation algorithms.

In the analysis that is to follow, we will make use of the vectorised
version of~\eqref{eq:chan_model}. Let $\mat{y} = \vectorize(\mat{Y}) =
(Y_{1,1}, \ldots, Y_{L,1}, Y_{1,2}, \ldots, Y_{L,2}, \ldots, Y_{1,K},
\ldots, Y_{L,K})'$, and $\mat{z} = \vectorize(\mat{Z})$ then
\begin{equation}
  \mat{y} = \mat{\Omega}(\mat{\tau},\mat{\nu},\mat{X}) \mat{a} + \mat{z}. \label{eq:vec_model}
\end{equation}
where the $KL \times P$ matrix $\mat{\Omega}$ is a function of
$\mat{\tau}$, $\mat{\nu}$, and $\mat{X}$ as follows,
\begin{equation} \label{eq:omega_mat}
 \mat{\Omega}(\mat{\tau},\mat{\nu},\mat{X}) = \left(
\begin{matrix}
X_{1,1} \psi_{1}(\nu_1) \phi^*_1(\tau_1) & X_{1,1} \psi_{1}(\nu_2) \phi^*_1(\tau_2) & \ldots  & X_{1,1} \psi_{1}(\nu_P) \phi^*_1(\tau_P) \\
\vdots & \vdots & & \vdots \\
X_{L,1} \psi_{L}(\nu_1) \phi^*_1(\tau_1) & X_{L,1} \psi_{1}(\nu_2) \phi^*_1(\tau_2) & \ldots  & X_{L,1} \psi_{L}(\nu_P) \phi^*_1(\tau_P) \\
X_{1,2} \psi_{1}(\nu_1) \phi^*_2(\tau_1) & X_{1,2} \psi_{1}(\nu_2) \phi^*_2(\tau_2) & \ldots  & X_{1,2} \psi_{1}(\nu_P) \phi^*_2(\tau_P) \\
\vdots & \vdots & & \vdots \\
X_{L,2} \psi_{L}(\nu_1) \phi^*_2(\tau_1) & X_{L,2} \psi_{1}(\nu_2) \phi^*_2(\tau_2) & \ldots  & X_{L,2} \psi_{L}(\nu_P) \phi^*_2(\tau_P) \\
\vdots & \vdots & & \vdots \\
X_{1,K} \psi_{1}(\nu_1) \phi^*_K(\tau_1) & X_{1,K} \psi_{1}(\nu_2) \phi^*_K(\tau_2) & \ldots  & X_{1,K} \psi_{1}(\nu_P) \phi^*_K(\tau_P) \\
\vdots & \vdots & & \vdots \\
X_{L,K} \psi_{L}(\nu_1) \phi^*_K(\tau_1) & X_{L,K} \psi_{L}(\nu_2) \phi^*_K(\tau_2) & \ldots  & X_{L,K} \psi_{L}(\nu_P) \phi^*_K(\tau_P)
\end{matrix}
\right),
\end{equation}
where $(\cdot)^*$ denotes the complex conjugate.

%
%
\section{Ambiguity Function} \label{sec:ambiguity}

The ambiguity function of the transmitted signal $x(t)$ is defined as
the inner product of the signal with a delayed, frequency shifted
version of itself~\cite{Skolnik:2002book}
\begin{equation} \label{eq:ambig_def}
A_x(\tau,\nu) = \int_{-\infty}^{\infty} x(t) x^*(t-\tau) e^{-j 2 \pi \nu t} \, d t.
\end{equation}
In the context of OFDM communications the ambiguity function has been
used often as a tool for pulse design and
optimisation~\cite{Kozek:1998p1579,Liu:2004p2583}. In radar systems,
the ambiguity function plays an important role in determining target
range and velocity resolution~\cite{Skolnik:2002book}. In this section
we derive the ambiguity function of the transmitted OFDM signal and
highlight important characteristics that will affect the delay/Doppler
estimation problem. Moreover, as we shall see later, parts of the
estimation problem can be written succinctly in terms of the ambiguity
function. In this direction, substitution of~\eqref{eq:tx_ofdm_sym}
into~\eqref{eq:ambig_def} yields the following result.
\begin{theorem}[OFDM Ambiguity Function]
  For a general windowing function $w(t)$, let $A_w(\nu,\tau)$ denote
  its ambiguity function. The ambiguity function of the OFDM
  signal~\eqref{eq:tx_ofdm_sym} is,
\begin{align}
A_x(\tau,\nu) = \frac{e^{-j \pi K \tau/T}}{KL} \sum_{l,k,l',k'} &X_{l,k} X^*_{l',k'} e^{-j 2 \pi (k-k')T_{\rm cp}/T} e^{j 2 \pi (k'-1) \tau/T} e^{-j 2 \pi T_d (l-1) \left(\nu - \frac{k-k'}{T}\right)} \notag \\
& \times A_{w}(\tau + (l'-l)T_d, \nu + (k'-k)/T). \label{eq:ambig_fun}
\end{align}
\end{theorem}
Due to the quadruple summation, numerical evaluation
of~\eqref{eq:ambig_fun} is computationally demanding. However, under
some common practical design assumptions we can make further
simplifications. Firstly, the windowing function is usually designed
such that $A_w(\tau,\nu) \approx 0$ for $|\tau| > T_{d}$. Window
functions of the form~\eqref{eq:window_fun} will have this
property. Thus the terms when $l' \neq l$ in the summation
of~\eqref{eq:ambig_fun} are approximately zero. Secondly, although
$A_w(\tau,\nu)$ cannot be considered negligible for $|\nu| >
(k'-k)/T$, since $X_{l,k}$ are i.i.d. with zero mean and unit variance
by assumption (except for the null subcarriers, which have zero
power), the summation of terms over $k \neq k'$ will approach zero for
large $K$ and/or $L$. In addition, since we are primarily concerned
with delay and Doppler in the region $ 0 \leq \tau \leq T_{\rm cp}$
and $|\nu| \ll 1/T$, where there is negligible variation in
$A_w(\tau,\nu)$, we may assume $A_w(\tau,\nu) \approx A_w(0,0)$, which
only introduces a constant phase offset, since $|A_w(0,0)|^2 =
1$. Hence we ignore the complex scaling affects of $A_w(\tau,\nu)$ and
approximate~\eqref{eq:ambig_fun} as
\begin{align} \label{eq:ambig_fun_approx}
A_x(\tau,\nu) \approx \tilde{A}_x(\tau,\nu) \triangleq \frac{1}{KL} \sum_{k,l} |X_{l,k}|^2 e^{-j 2 \pi (l-1) \nu T_d} e^{j 2 \pi (k-1-\lfloor K/2 \rfloor) \tau/T}.
\end{align}
In light of~\eqref{eq:ambig_fun}, note
that~\eqref{eq:ambig_fun_approx} is also the ambiguity function when
ISI and ICI can be considered negligible.

\begin{figure}[htbp]
  \centering 
  \subfigure[$K=52$, $L=90$]{\includegraphics[width=0.47\columnwidth]{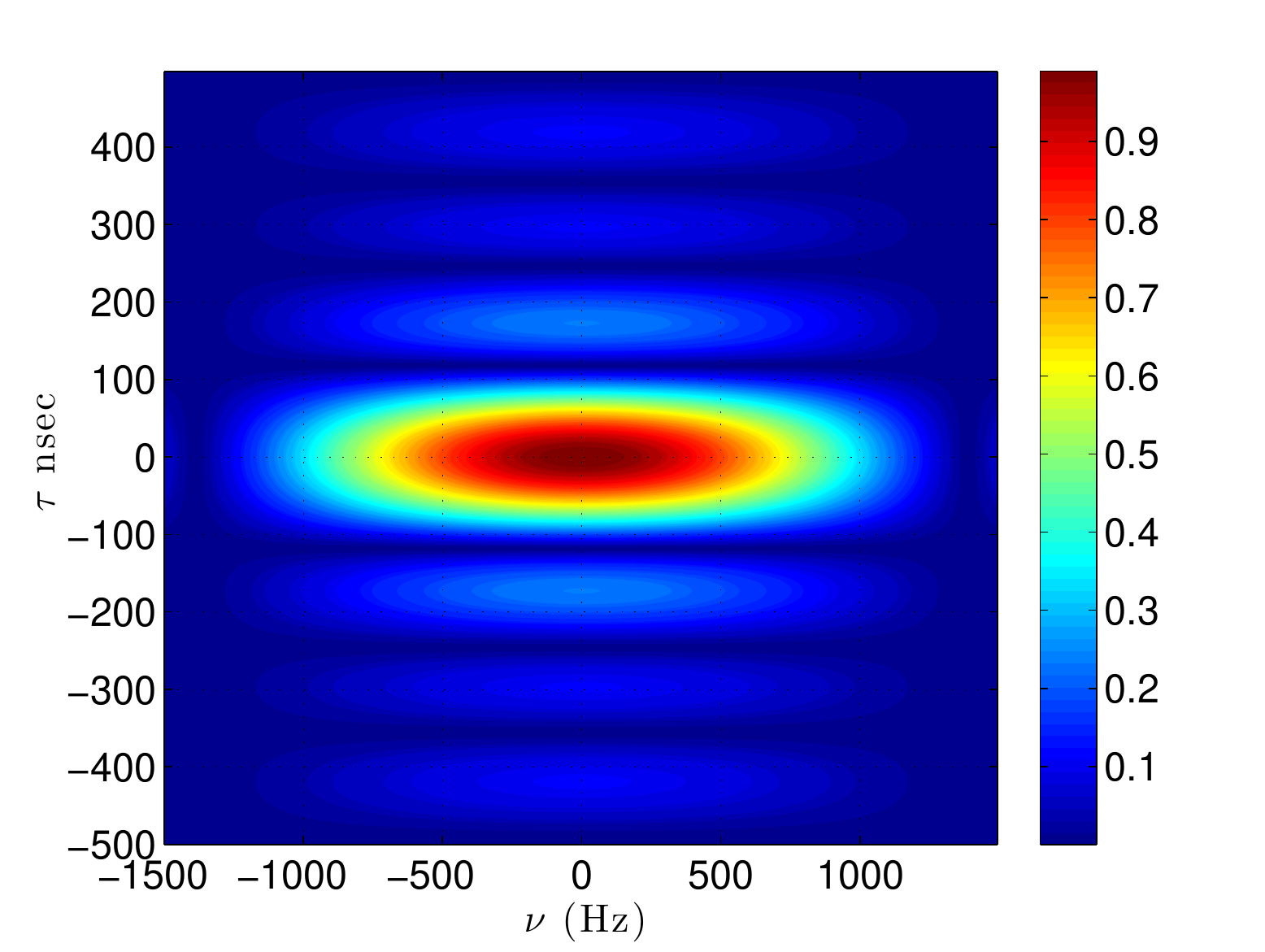} \label{fig:ambig_K52_L90}}
  \subfigure[$K=52, L=242$]{\includegraphics[width=0.47\columnwidth]{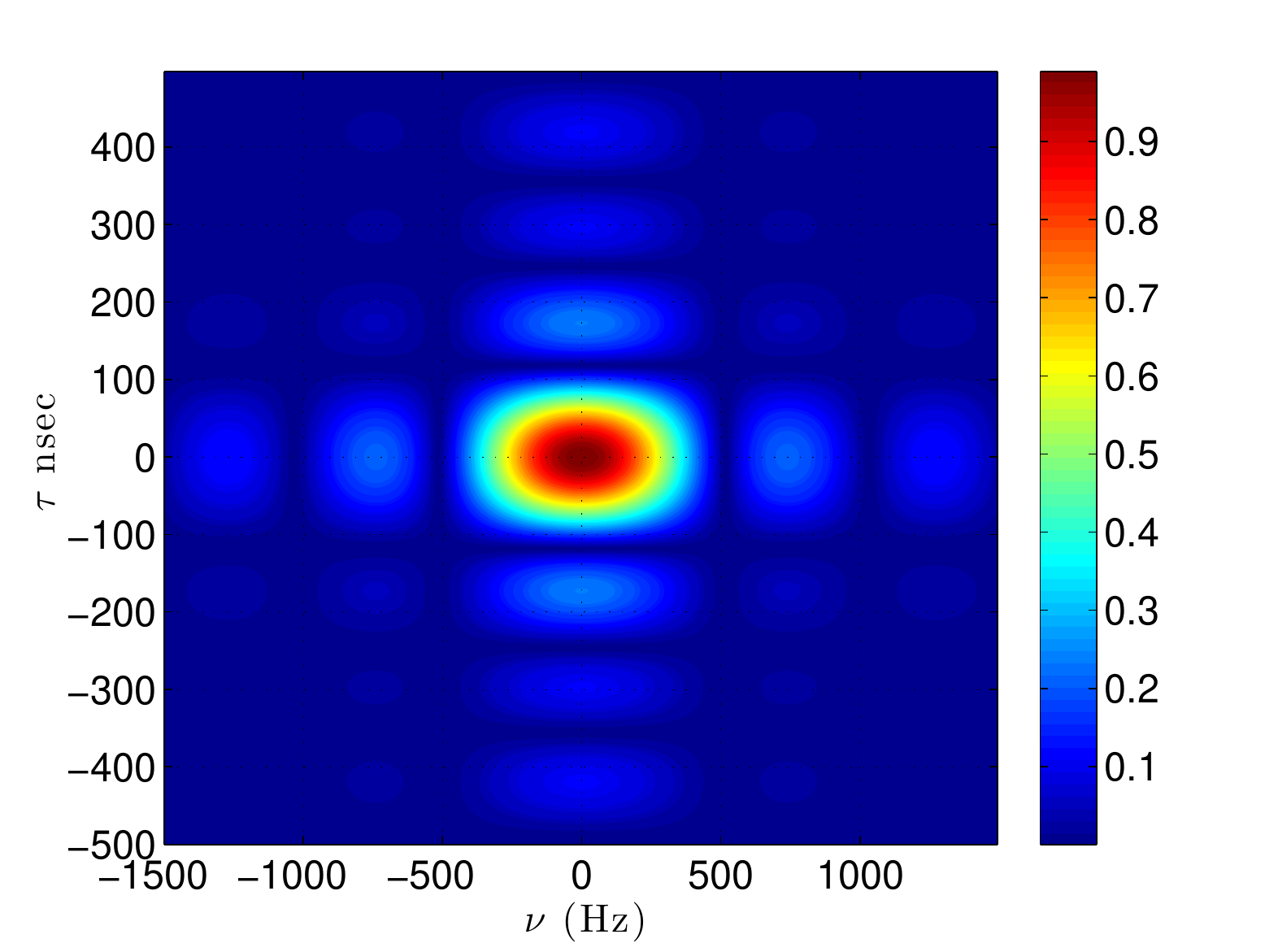}   \label{fig:ambig_K52_L242}}
  \caption{Magnitude contour plot of the ambiguity
    function~\eqref{eq:ofdm_ambig_psk_zero_dc} of an 802.11a OFDM
    system with PSK modulation, $T = 6.4$ $\mu$sec, $T_d = 8$
    $\mu$sec}
\label{fig:ambig}
\end{figure}

For phase shift keying (PSK) modulation, $|X_{l,k}| = 1$ with no null
subcarriers, i.e. $\mathcal{K} = \emptyset$, then using the geometric
summation formula, the ambiguity function~\eqref{eq:ambig_fun_approx}
further simplifies to
\begin{equation}
\tilde{A}_x(\tau,\nu) \approx e^{-j \pi K \tau/T} \sinc \left( \pi K \frac{\tau}{T} \right) \sinc \left( \pi L \nu T_d \right),
 \label{eq:ofdm_ambig_psk}
\end{equation}
where $\sinc(x) = \sin(x)/x$. Note that~\eqref{eq:ofdm_ambig_psk} is
also equal to the expectation of~\eqref{eq:ambig_fun} over the
transmitted symbols $X_{l,k}$, i.e. it is the expected ambiguity
function, $\EE \left[ A_x(\tau,\nu) \right]$, for an arbitrary signal
constellation $\mathcal{X}$ with zero mean and unit variance. In many
OFDM standards, subcarrier $k = \lfloor
K/2 \rfloor +1$ is a null subcarrier. For these special
cases~\eqref{eq:ofdm_ambig_psk} becomes,
\begin{equation}
\tilde{A}_x(\tau,\nu) \approx e^{-j \pi K \tau/T} \cos \left[ \pi (K/2 + 1)(\tau/T) \right] \sinc ( \pi \tau K/(2T)) \sinc (\pi \nu T_d L)   \label{eq:ofdm_ambig_psk_zero_dc}
\end{equation}
From the above expressions~\eqref{eq:ofdm_ambig_psk}
and~\eqref{eq:ofdm_ambig_psk_zero_dc}, we see that the $\sinc$ terms
introduce sidelobes in the ambiguity function. Interestingly, the
sidelobes are de-coupled in time and frequency.  As an example, the
ambiguity function of the 802.11a standard~\cite{ieee802.11a,OhaPet99} is plotted
in Fig.~\ref{fig:ambig}, i.e. $K=53$ subcarriers, with a null
subcarrier at $k = \lfloor K/2 \rfloor + 1$, $T_d = 8$ $\mu$sec and $T
= 6.4$ $\mu$sec.
From Fig.~\ref{fig:ambig}, as predicted by
\eqref{eq:ofdm_ambig_psk_zero_dc} we see that increasing $L$ improves
the Doppler resolution, but not the delay resolution, which is
dependent only on $K$ and the subcarrier spacing ($1/T$).

%
%
\section{Multipath Parameter Estimation} \label{sec:estimator}

Our primary objective is to estimate $\mat{a} = (a_1,\ldots,a_P)$,
$\mat{\tau} = (\tau_1,\ldots,\tau_P) $ and $\mat{\nu} =
(\nu_1,\ldots,\nu_P)$ in~\eqref{eq:cir} from the received noisy
symbols $\mat{Y}$~\eqref{eq:matched_filt_out} given perfect knowledge
of $\mat{X}$. In this section, without loss of generality, for brevity
of notation, we assume the OFDM system has no null subcarriers,
i.e. $\mathcal{K} = \emptyset$.  Using~\eqref{eq:vec_model}, the
maximum likelihood (ML) approach is to solve the following
\begin{equation}
(\mat{\hat{a}}, \mat{\hat{\tau}}, \mat{\hat{\nu}}) = \arg \min_{\mat{a},\mat{\tau},\mat{\nu}} \| \mat{y} -  \mat{\Omega}(\mat{\tau},\mat{\nu},\mat{X}) \mat{a} \|^2, \label{eq:ml_problem}
\end{equation}
which is a non-linear least squares minimisation problem. The
computational complexity can be reduced by replacing $\mat{a}$ with
its least squares estimate. That is, for a given $\mat{\tau}$ and
$\mat{\nu}$ the ML estimate of $\mat{a}$ is a linear least squares
minimisation problem, which has solution~\cite{Boyd:2004book}
\begin{equation}
\mat{\hat{a}} = \left( \mat{\Omega}^{\dagger}\mat{\Omega} \right)^{-1} \mat{\Omega}^{\dagger} \mat{y}, \label{eq:lse_a}
\end{equation}
where we have dropped the dependence of $\mat{X},\mat{\tau}$ and
$\mat{\nu}$ for brevity of notation. Hence
substituting~\eqref{eq:lse_a} for $\mat{a}$ in~\eqref{eq:ml_problem}
results in the reduced problem
\begin{equation}
(\mat{\hat{\tau}}, \mat{\hat{\nu}}) = \arg \max_{\mat{\tau},\mat{\nu}} \; \mat{y}^{\dagger} \mat{\Omega} (\mat{\Omega}^{\dagger} \mat{\Omega})^{-1} \mat{\Omega}^{\dagger} \mat{y}. \label{eq:reduced_ml_problem2}
\end{equation}
It is known that problems~\eqref{eq:reduced_ml_problem2}
and~\eqref{eq:ml_problem} are equivalent,
i.e.~\eqref{eq:reduced_ml_problem2} followed by~\eqref{eq:lse_a} is
also the ML
solution~\cite{Golub:1973p413,Bresler:1986p1081}. Unfortunately,~\eqref{eq:reduced_ml_problem2}
is in general multimodal, rendering the multidimensional search for a
global extremum computationally prohibitive.

Before we begin our reduced complexity suboptimal solution, let us
first make some interesting observations
about~\eqref{eq:reduced_ml_problem2}. Let $\mat{R} =
\mat{\Omega}^{\dagger} \mat{\Omega}$ and $\mat{w} =
\mat{\Omega}^{\dagger} \mat{y}$. From~\eqref{eq:omega_mat}, it is
straightforward to show,
\begin{align}
R_{ij} &= KL \tilde{A}_x(\tau_i-\tau_j, \nu_j-\nu_i) \label{eq:mat_R} \\
w_i &= \mat{\psi}^{\dagger}(\nu_i) \left( \mat{Y} \odot \mat{X}^* \right) \mat{\phi}(\tau_i), \label{eq:vec_w}
\end{align}
for $i,j = 1,\ldots,P$, where $\mat{\psi}(\nu_i)$ and
$\mat{\phi}(\tau_i)$ denote column $i$ of the matrices
$\mat{\Psi}$ and $\mat{\Phi}$ respectively, and $\mat{A}^*$ denotes
the element-wise conjugate of the matrix $\mat{A}$. Thus, rather than
performing computation of $\mat{R} = \mat{\Omega}^{\dagger}
\mat{\Omega}$ (requiring on the order of $PKL(KL+1)/2$ complex
multiply-accumulate operations) using standard matrix operations, to
reduce complexity, $\mat{R}$ can be evaluated using the ambiguity
function via a look-up table. Moreover, for the special case of PSK
modulation,~\eqref{eq:ofdm_ambig_psk} implies we only need the
evaluation of a $\sinc(x)$ function.


For the special case of $P=1$, the ML
solution~\eqref{eq:reduced_ml_problem2} becomes
\begin{align}
(\hat{\tau}_1, \hat{\nu}_1) &= \arg \max_{\tau,\nu} \; \left|  \mat{\psi}^{\dagger}(\nu_1) \left( \mat{Y} \odot \mat{X}^* \right) \mat{\phi}(\tau_1) \right|^2,   \label{eq:ml_single_tap}
\end{align}
after which the corresponding complex gain ML estimates can be
determined using~\eqref{eq:lse_a}, 
\begin{equation}
\hat{a}_1 = \frac{1}{KL}   \mat{\psi}^{\dagger}(\hat{\nu}_1) \left( \mat{Y} \odot \mat{X}^* \right) \mat{\phi}(\hat{\tau}_1).
\end{equation}
We see that the solution to~\eqref{eq:reduced_ml_problem2} corresponds
to the maximum absolute value of the 2-D
periodogram~\cite{Kay:1988book}. Moreover, the CRLB for the estimation
of a single tap multipath channel can be written
as~\cite{Chien:1981thesis}
\begin{align}
\mathrm{var} [\hat{\nu}_1 T_d] \geq  \frac{1}{4 \pi^2} \frac{6}{KL(L^2 - 1)} \frac{\sigma^2}{|a_1|^2}, \;\;\;\;\;\;
\mathrm{var} \left[ \hat{\tau}_1/T \right] \geq  \frac{1}{4 \pi^2}\frac{6}{KL(K^2 - 1)} \frac{\sigma^2}{|a_1|^2}. \label{eq:crlb_single_tap}
\end{align}
Note that Kay and Nekovei~\cite{Kay:1990p1807} proposed a low
complexity weighted phase averager estimator as an alternative to
solving~\eqref{eq:ml_single_tap}.

If we were to use~\eqref{eq:ml_single_tap} when multiple taps are
present ($P>1$), then
\begin{equation}
 \mat{\psi}^{\dagger}(\nu) \left( \mat{Y} \odot \mat{X}^* \right) \mat{\phi}(\tau) = KL \sum_p a_p \tilde{A}_x(\nu_p-\nu,\tau - \tau_p) + \mat{\psi}^{\dagger}(\nu) \left[ \mat{Z} \odot \mat{X}^*\right] \mat{\phi}(\tau), \notag
\end{equation}
which is the superimposition of complex scaled, delay and frequency
shifted ambiguity functions, plus an additive Gaussian noise term. We
see that detection and estimation of a particular tap will be
significantly affected by the main lobe and sidelobes from the
ambiguity functions of the remaining taps. This motivates a successive
cancellation approach whereby the signal contribution in $\mat{Y}$
induced by a multipath tap is removed after it is detected, thus
allowing subsequent taps to be detected and estimated. Successive
cancellation algorithms have found widespread use in a number of
communication scenarios requiring the recovery of multiple
superimposed signals. In particular, interference cancellation
(successive and parallel forms) is the basis of practical
low-complexity multi-user decoding algorithms, which attain close to
single-user bit error rate performance~\cite{Schlegel:2006book}. In
our case, the superimposed signals are not signals from multiple
users, but time/frequency shifted versions of the same signal. However
the same principle can still be applied, and as we will see later,
achieves performance close to the CRLB of a single tap channel
(provided the taps are sufficiently separated in either delay or
Doppler). In this direction, the first algorithm we propose is based
on successive cancellation and is employed to find an initial estimate
of the delay, Doppler and complex gain of each tap. The second
algorithm we propose is based on parallel cancellation and is employed
to refine the initial estimates. Integral to both of these algorithms
is a search for the largest absolute value of a 2-D
periodogram~\cite{Kay:1988book}, and we propose a low-complexity 2-D
bisection algorithm for doing this. A detailed description of each of
these algorithms is given as follows.

\subsection{Initial Estimation}

Algorithm~\ref{alg:init_sc} describes our proposed initial successive
cancellation procedure. First we initialise the residual error matrix
$\mat{E}^{(1)}$ equal to the received noisy OFDM symbols $\mat{Y}$. At
iteration $p=1,2,\dots,P$: we find $\hat{\tau}_p$ and $\hat{\nu}_p$
that correspond to the maximum absolute value squared of the 2-D
periodigram of $\mat{E}^{(p)}$; construct the $p \times p$ matrix
\begin{equation*}\mat{R}^{(p)} =
\begin{pmatrix}
  R_{11} & R_{12} &\dots & R_{1p} \\
  \vdots & \ddots & & \\
    R_{p1} & R_{p2} &\dots & R_{pp} 
\end{pmatrix}
\end{equation*}
and length $p$ vector $\mat{w}^{(p)}=(w_1,w_2,\dots,w_p)$ substituting
the delay and Doppler estimates
$\hat{\mat{\tau}}^{(p)}=(\hat{\tau}_1,\ldots, \hat{\tau}_p)$ and
$\hat{\mat{\nu}}^{(p)}=(\hat{\nu}_1, \ldots, \hat{\nu}_p)$
into~\eqref{eq:mat_R} and~\eqref{eq:vec_w}; re-estimate the length $p$
complex gain vector $\hat{\mat{a}}^{(p)} = (\hat{a}_1 \ldots,
\hat{a}_p) = (\mat{R}^{(p)})^{-1} \mat{w}^{(p)}$; and finally subtract
the signal contributions of all $p$ estimated multipath components
from $\mat{Y}$, which becomes the residual error matrix for the next
iteration. 


\begin{algoendfloat}
\caption{Initial estimation via successive cancellation.}
\label{alg:init_sc}
\begin{algorithmic}[1]
\STATE $\mat{E}^{(1)} = \mat{Y}$ \label{alg:init_sc:init}
\FOR{$p = 1,\ldots,P$}
\STATE $(\hat{\tau}_p, \hat{\nu}_p) = \arg \max_{\tau,\nu} \; \left|  \mat{\psi}^{\dagger}(\nu) \left( \mat{E}^{(p)} \odot \mat{X}^* \right) \mat{\phi}(\tau) \right|^2 $  \label{alg:init_sc:line:periodogram}
\STATE Contruct $\mat{R}^{(p)}$ and $\mat{w}^{(p)}$ using \eqref{eq:mat_R} and \eqref{eq:vec_w} with $\hat{\tau}_1,\ldots,\hat{\tau}_p$ and $\hat{\nu}_1,\ldots, \hat{\nu}_p$.
\STATE $\mat{\hat{a}}^{(p)} = (\mat{R}^{(p)})^{-1} \mat{w}^{(p)} $
\STATE $\mat{E}^{(p+1)} = \mat{Y} - \left[ \mat{\Psi}(\mat{\hat{\nu}}^{(p)}) \diag(\mat{\hat{a}}^{(p)}) \mat{\Phi}^{\dagger}(\mat{\hat{\tau}}^{(p)}) \right] \odot \mat{X} $ \label{alg:init_sc:line:subtract}
\ENDFOR
\end{algorithmic}
\end{algoendfloat}
Typically Algorithm~\ref{alg:init_sc} will estimate the multipath
starting from the strongest to the weakest tap, i.e.  $|\hat{a}_1| >
|\hat{a}_2| > \ldots > |\hat{a}_P |$. Thus, for the case when $P$ is
unknown, an obvious exit criterion is to stop once $|\hat{a}_p| <
\gamma$, where $\gamma$ is a threshold that determines the minimum tap
energy. Alternatively, the Algorithm can be modified to incorporate a
model order selection rule~\cite{Stoica:2004p36}.

Note that two simple modifications can be made to
Algorithm~\ref{alg:init_sc} to further reduce complexity. Firstly, in
the main loop, rather than subtracting all multipath contributions of
the previously estimated components from the original signal $\mat{Y}$
to obtain the residual error $\mat{E}^{(p)}$, simply subtract the
contribution of the current estimate from the residual error of the
previous iteration $\mat{E}^{(p-1)}$,
i.e. line~\ref{alg:init_sc:line:subtract} can be replaced with
$\mat{E}^{(p)} = \mat{E}^{(p-1)} - \left[ \hat{a}^{(p)}_{p}
  \mat{\psi}(\hat{\nu}_{p}) \mat{\phi}^{\dagger}(\hat{\tau}_p) \right]
\odot \mat{X} $. Secondly, rather than operating on $\mat{Y}$, one
could apply the algorithm on the zero-forcing estimate of $\mat{H}$,
i.e. $\hat{H}_{l,k} = Y_{l,k} X^{*}_{l,k}/ |X_{l,k}|$. Thus, in
Algorithm~\ref{alg:init_sc}, one simply replaces $\mat{Y}$ with
$\mat{\hat{H}}$ and the Hadamard product with $\mat{X}$ in
lines~\ref{alg:init_sc:line:periodogram}
and~\ref{alg:init_sc:line:subtract} is no longer required. To
summarize, we can make the following complexity-reducing modifications to
Algorithm~\ref{alg:init_sc}.  Line~\ref{alg:init_sc:init}:
$\mat{E}^{(1)} = \hat{\mat{H}}$,
Line~\ref{alg:init_sc:line:periodogram}: $(\hat{\tau}_p, \hat{\nu}_p)
= \arg \max_{\tau,\nu} \; \left| \mat{\psi}^{\dagger}(\nu)
  \mat{E}^{(p)} \mat{\phi}(\tau) \right|^2 $, and
Line~\ref{alg:init_sc:line:subtract}: $\mat{E}^{(p+1)} = \hat{\mat{H}}
- \left[ \mat{\Psi}(\mat{\hat{\nu}}^{(p)}) \diag(\mat{\hat{a}}^{(p)})
  \mat{\Phi}^{\dagger}(\mat{\hat{\tau}}^{(p)}) \right]$

\subsection{Estimation Refinement}
It is quite reasonable to rely solely on Algorithm~\ref{alg:init_sc}
to estimate the delay/Doppler. Indeed similar approaches have been
employed in~\cite{Thomas:2003p74,Kliger:2005p2563,Berger:2008p2384},
but without any detailed comparison to theoretical bounds. We find
that the performance of Algorithm~\ref{alg:init_sc} is hampered by
interference from undetected taps, which as we will see later,
introduces a floor in the root mean squared (RMS) error
performance. Therefore we propose a refinement process based on
parallel cancellation whereby for each iteration, all multipath
components are removed except for the component of interest, that is
subsequently re-estimated. This refinement procedure is described in
detail in Algorithm~\ref{alg:refine}, where $\mat{\hat{\tau}}^{(i)} =
(\hat{\tau}_1^{(i)}, \ldots, \hat{\tau}_P^{(i)}$),
$\mat{\hat{\nu}}^{(i)} = (\hat{\nu}_1^{(i)}, \ldots,
\hat{\nu}_P^{(i)})$ and $\mat{\hat{a}}^{(i)} =
(\hat{a}_1^{(i)},\ldots,\hat{a}_P^{(i)})$ denote the refined estimates
after the $i$'th iteration, and $\mat{\hat{\tau}}^{(0)} =
\mat{\hat{\tau}}$, $\mat{\hat{\nu}}^{(0)} = \mat{\hat{\nu}}$ and
$\mat{\hat{a}}^{(0)} = \mat{\hat{a}}$ are the initial estimates
obtained from Algorithm~\ref{alg:init_sc}. In addition, we let
$\mat{\hat{\tau}}^{(i)}_{\bar{p}}$, $\mat{\hat{\nu}}^{(i)}_{\bar{p}}$
and $\mat{\hat{a}}^{(i)}_{\bar{p}}$ denote the  refined
estimates at step $i$ with element $p$ element omitted.
\begin{algoendfloat}
\caption{Estimate refinement algorithm.}
\label{alg:refine}
\begin{algorithmic}[1]
  \STATE $\mat{\hat{\tau}}^{(0)} = \mat{\hat{\tau}}$,$\mat{\hat{\nu}}^{(0)} = \mat{\hat{\nu}}$ and $\mat{\hat{a}}^{(0)} = \mat{\hat{a}}$
  \FOR{$i = 1,\ldots,N$}
  \FOR{$p=1,\ldots,P$}
  \STATE $\mat{E} = \mat{Y} - \left[ \mat{\Psi}(\mat{\hat{\nu}}^{(i-1)}_{\bar{p}}) \diag(\mat{\hat{a}}^{(i-1)}_{\bar{p}}) \mat{\Phi}^{\dagger}(\mat{\hat{\tau}}^{(i-1)}_{\bar{p}}) \right] \odot \mat{X} $  \label{alg:refine:line:subtract}
  \STATE $(\hat{\tau}^{(i)}_q, \hat{\nu}^{(i)}_q) = \arg \max_{\tau,\nu} \; \left|  \mat{\psi}^{\dagger}(\nu) \left( \mat{E} \odot \mat{X}^* \right) \mat{\phi}(\tau) \right|^2 $ \label{alg:refine:line:periodogram}
  \ENDFOR
  \STATE $\hat{a}^{(i)} = \mat{R}^{-1}(\mat{\hat{\tau}}^{(i)},\mat{\hat{\nu}}^{(i)}) \mat{w}(\mat{\hat{\tau}}^{(i)},\mat{\hat{\nu}}^{(i)})$
\ENDFOR
\end{algorithmic}
\end{algoendfloat}

Note that rather than refining for a fixed number of iterations,
Algorithm~\ref{alg:refine} can be easily be modified to incorporated
an early stopping criterion, e.g. by checking the improvement in the
residual error $\| \mat{E}\|^2$. As previously described for Algorithm~\ref{alg:init_sc}, one could
apply Algorithm~\ref{alg:refine} to the zero-forcing estimate of
$\mat{\hat{H}}$, i.e. replace $\mat{Y}$ with $\mat{\hat{H}}$ and
removing the Hadamard product with $\mat{X}$ in
lines~\ref{alg:refine:line:subtract}
and~\ref{alg:refine:line:periodogram}.

\subsection{2-D Bisection Algorithm}

As mentioned earlier, the maximisation step in line
\ref{alg:init_sc:line:periodogram} Algorithm \ref{alg:init_sc} and
line \ref{alg:refine:line:periodogram} of Algorithm~\ref{alg:refine}
can be solved by finding the maximum absolute value of the 2-D
periodogram~\cite{Kay:1988book}. To perform this operation, we propose
a 2-D bisection approach described as follows. First we assume $ \tau_p
\in (\tau_{\rm min}, \tau_{\rm max})$ and $ \nu_p \in(\nu_{\rm min},
\nu_{\rm max})$ for all $p = 1,\ldots,P$, i.e. the delay/Doppler of
each tap is constrained to lie within predefined intervals. Let
$(\tau^{(i)}_{\rm min}, \tau^{(i)}_{\rm max})$ and $ (\nu^{(i)}_{\rm
  min}, \nu^{(i)}_{\rm max})$ denote the search interval at iteration
$i$, and $\mat{\tilde{\tau}}^{(i)}$ and $\mat{\tilde{\nu}}^{(i)}$
denote linearly spaced vectors within these intervals, i.e.
\begin{equation}
\tilde{\tau}_{m}^{(i)} = \tau_{\rm min}^{(i)} + (m-1)\Delta \tau^{(i)} \;\;\;\;\;
\tilde{\nu}_{m}^{(i)} = \nu_{\rm min}^{(i)} + (n-1)\Delta \nu^{(i)}, \label{eq:lin_space}
\end{equation}
for $m = 1,\ldots,M$ and $n = 1,\ldots,N$, where $\Delta \tau^{(i)} =
(\tau_{\rm max}^{(i)} - \tau_{\rm min}^{(i)})/M$ and $\Delta \nu^{(i)}
= (\nu_{\rm max}^{(i)} - \nu_{\rm min}^{(i)})/N$ denote the bin
spacing at the $i$'th iteration. For each iteration of the bisection
algorithm, we find the indices corresponding to the largest peak of $
\mat{\Psi}^{\dagger}(\mat{\tilde{\nu}}^{(i)}) \left[ \mat{Y} \odot
  \mat{X}^* \right] \mat{\Phi}( \mat{\tilde{\tau}}^{(i)})$. For the
next iteration, the search interval is then bisected or reduced to a
smaller 2-D region, i.e. $\left( 2 \beta \Delta \tau^{(i)}, 2 \beta
  \Delta \nu^{(i)} \right)$, centered at the previous delay/Doppler
indices (typically $\beta \geq 1/2$). A detailed description of the
procedure is given in Algorithm~\ref{alg:bisection}. Note that for
ease of exposition, the bisection process completes after a fixed
number of iterations $N_{\rm bisect}$. The algorithm can easily be
modified to employ an early stopping criterion, e.g. exit the main
loop when $\Delta \tau^{(i)} < \epsilon_{\tau}$ and $\Delta \nu^{(i)}
< \epsilon_{\nu}$ to ensure a certain level of delay/Doppler
resolution.
\begin{algoendfloat}  
\caption{2-D Bisection Algorithm.}
\label{alg:bisection}
\begin{algorithmic}[1]
  \STATE Initialise $ (\tau_{\rm min}^{(0)}, \tau_{\rm max}^{(0)}) =
  (\tau_{\rm min}, \tau_{\rm max})$ and $(\nu_{\rm min}^{(0)},
  \nu_{\rm max}^{(0)}) = (\nu_{\rm min}, \nu_{\rm max})$.
  \FOR{ $i = 1,\ldots, N_{\rm bisect}$}
  \STATE $\Delta \tau^{(i)} = (\tau_{\rm max}^{(i-1)} - \tau_{\rm min}^{(i-1)})/M$, $\Delta \nu^{(i)}= (\nu_{\rm max}^{(i-1)} - \nu_{\rm min}^{(i-1)})/N$  
  \STATE Construct $\mat{\tilde{\tau}}^{(i)}$ and $\mat{\tilde{\nu}}^{(i)}$ using \eqref{eq:lin_space}.
  \STATE $\mat{\Upsilon}^{(i)} = \mat{\Psi}^{\dagger}(\mat{\tilde{\nu}}^{(i)}) \left[ \mat{Y} \odot \mat{X}^* \right] \mat{\Phi}( \mat{\tilde{\tau}}^{(i)})$ 
  \STATE $ (\hat{n}, \hat{m} ) = \arg \max_{n,m} |\Upsilon^{(i)}_{n,m}|^2$
  \STATE $\tau^{(i)}_{\rm max} = \tilde{\tau}^{(i)}_{\hat{m}} + \beta \Delta \tau^{(i)}$, $\tau^{(i)}_{\rm min} = \tilde{\tau}^{(i)}_{\hat{m}} - \beta \Delta \tau^{(i)}$
  \STATE $\nu^{(i)}_{\rm max} = \tilde{\nu}^{(i)}_{\hat{n}} + \beta \Delta \nu^{(i)}$, $\nu^{(i)}_{\rm min} = \tilde{\nu}^{(i)}_{\hat{n}} - \beta \Delta \nu^{(i)}$
  \ENDFOR
\STATE $\hat{\tau} = \tilde{\tau}^{(I)}_{\hat{m}}$, $\hat{\nu} = \tilde{\nu}^{(I)}_{\hat{n}}$.
\end{algorithmic}
\end{algoendfloat}

%
%
\section{Performance Evaluation} \label{sec:performance}

Performance evaluation is complicated by the fact there are infinitely
many possible multipath channel realisations and many OFDM system
design configurations all of which can have a significant effect on
the estimator's performance. To reduce our analysis, we focus on OFDM
systems with similar specifications to the IEEE802.11p standard (as
described in Section~\ref{sec:ambiguity}). In addition, we concentrate
on multipath channels typical of outdoor mobile vehicular
environments~\cite{Alexander:2007p108}, i.e. delay spreads not
exceeding $200$ nsec and Doppler differentials not exceeding $1000$
Hz. For example, at a carrier frequency of $5.9$ GHz, this corresponds
to a maximum excess delay of $60$ m and velocity
differentials of $51$ m/s or $183$ km/hr.

Ultimately, we would like to investigate the estimator's performance
for as many different multipath channel configurations as
possible. However, we find that the performance is significantly
affected by the location of the multipath taps in the 2-D
delay/Doppler space.  When two or more taps are too close to each
other there is a high probability Algorithm~\ref{alg:init_sc} will
detect these as a single tap.\footnote{In a physical sense, if these closely
  spaced taps are the result of first order reflections it may imply
  they are reflections from the same object.} The minimum separation
distance is essentially the delay/Doppler resolution of the estimator,
which is dependent on the main lobe of the ambiguity function, which
in turn, is dependent on the subcarrier spacing and duration of the
OFDM packet (as evidenced in~\eqref{eq:ofdm_ambig_psk}). When the
components are sufficiently separated, the estimator's performance is
dominated by AWGN and hence the CRLB~\eqref{eq:crlb_single_tap}.

To separate the above mentioned effects, we conducted Monte Carlo
simulations whereby for each trial a random set of multipath taps is
generated. Whilst these taps are drawn randomly, they are not i.i.d.,
and instead are drawn to ensure a minimum separation in delay and
Doppler. This is achieved by continually drawing a vector of $P$
delays from an i.i.d. uniform distribution on the interval $(\tau_{\rm
  min}, \tau_{\rm max})$ until the minimum pairwise distance between
the delays is greater than a specified $\Delta \tau$. The delays are
then sorted in ascending order. The Doppler offsets are generated in a
similar fashion on the interval $(\nu_{\rm min}, \nu_{\rm max})$, but
with no sorting. Note that $\Delta \tau \leq (\tau_{\max} - \tau_{\rm
  min})/P$ and similarly $\Delta \nu \leq (\nu_{\max} - \nu_{\rm
  min})/P$. Whilst we fix the power delay profile, for each trial, the
phase of each tap is generated randomly according to a uniform
distribution over the interval $(0,2\pi)$. Once the multipath taps are
generated, the frequency domain channel coefficients are generated
using~\eqref{eq:channel_coeffs} and the received noisy symbols are
generated using~\eqref{eq:chan_model}, where, without loss of
generality, we assume $X_{l,k} = 1$. It is important to note how the
error statistics were calculated. For each trial, RMS error statistics
were only collected when all taps are detected, i.e. each tap is
closest (in Euclidean distance) to a single estimate. Events when this
does not occur are counted as missed detections, but are not included in
the RMS error statistics. This allows us to separate error events
caused by miss detections due to the transmit ambiguity function.

In our simulations we considered a $P=3$ tap multipath channel, with
power delay profile $|a_1|^2 = 0$, $|a_2|^2 = -10$ and $|a_3|^2 = -20$
dB, $(\tau_{\rm min}, \tau_{\rm max}) = (0,200)$ nsec, $(\nu_{\rm
  min}, \nu_{\rm max}) = (-500,500)$, minimum delay separation of
$\Delta \tau = 66.67$ nsec and minimum Doppler separation of $\Delta
\nu = 333.33$ Hz.  Error statistics were collected from $10^4$
trials. 

\begin{figure}[htbp]
  \centering 
  \includegraphics[width=0.8\textwidth]{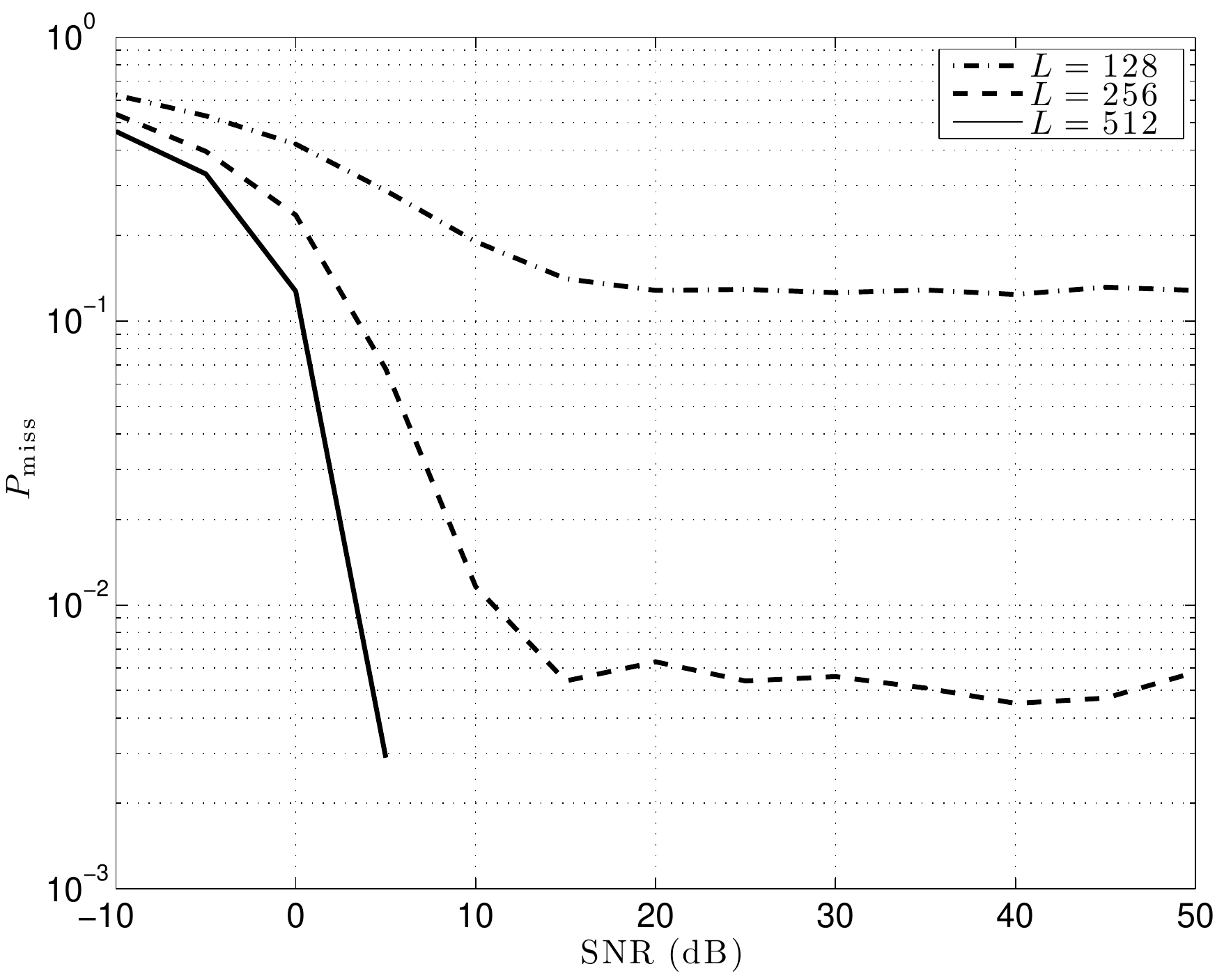}

  \caption{Probability of miss detecting all taps of a $P=3$ tap
    multipath channel.}
\label{fig:three_tap_miss_det}
\end{figure}

Fig.~\ref{fig:three_tap_miss_det} shows the miss detection
probability for $L = 128, 256$ and $512$ OFDM packet lengths.  We see
that when $L = 128$, the miss detection probability is greater than
$10$ percent. As $L$ increases the main lobe of the ambiguity function
shrinks in the Doppler domain improving the resolution of the
estimator and hence reduces the miss detection probability. When $L =
512$, no miss detections were recorded for an SNR greater than $5$ dB.

\begin{figure}[htbp]
  \centering 
  \subfigure[$L=128$]{\includegraphics[width=0.45\columnwidth]{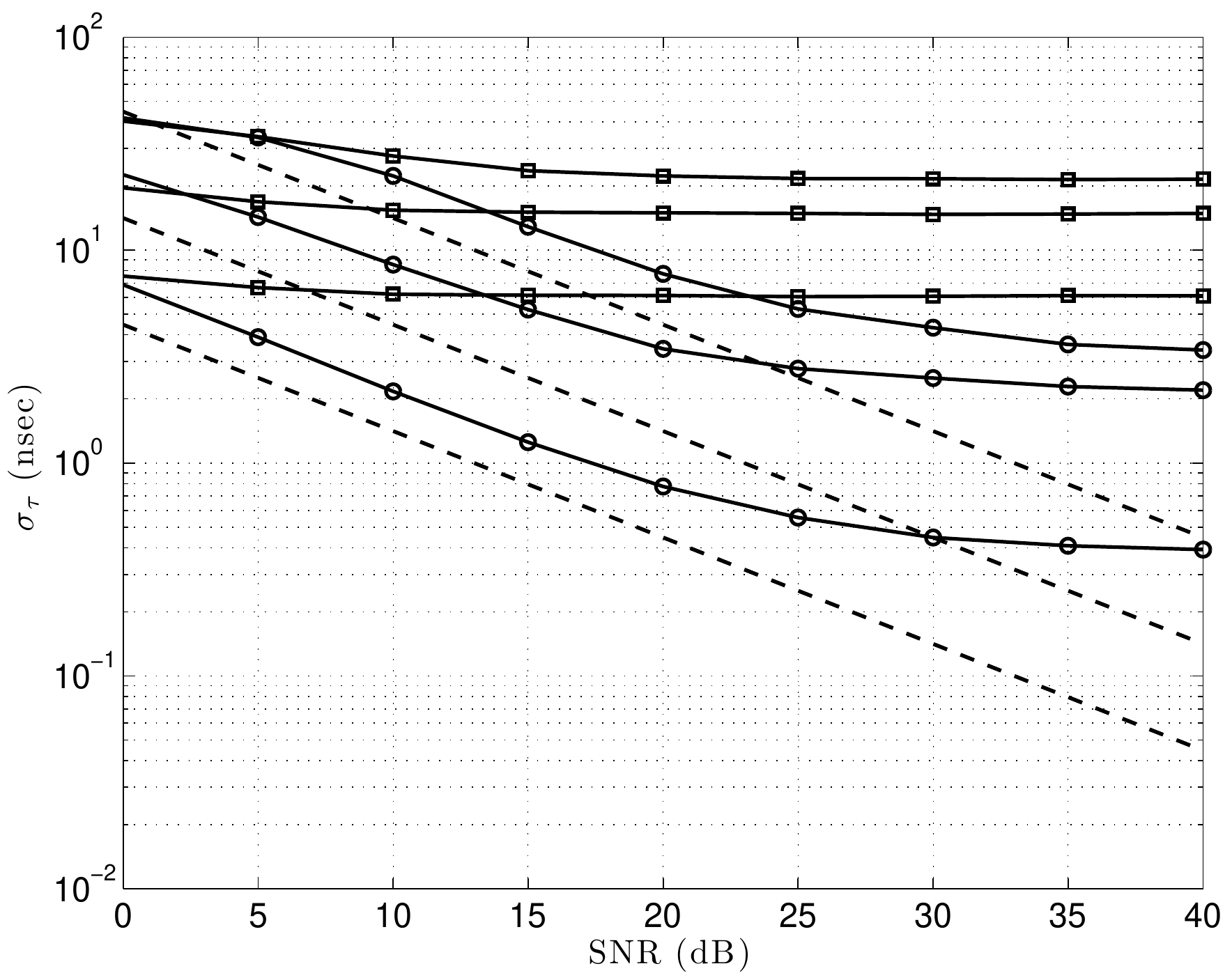} \label{fig:tau_L128_P3}}
  \subfigure[$L=128$]{\includegraphics[width=0.45\columnwidth]{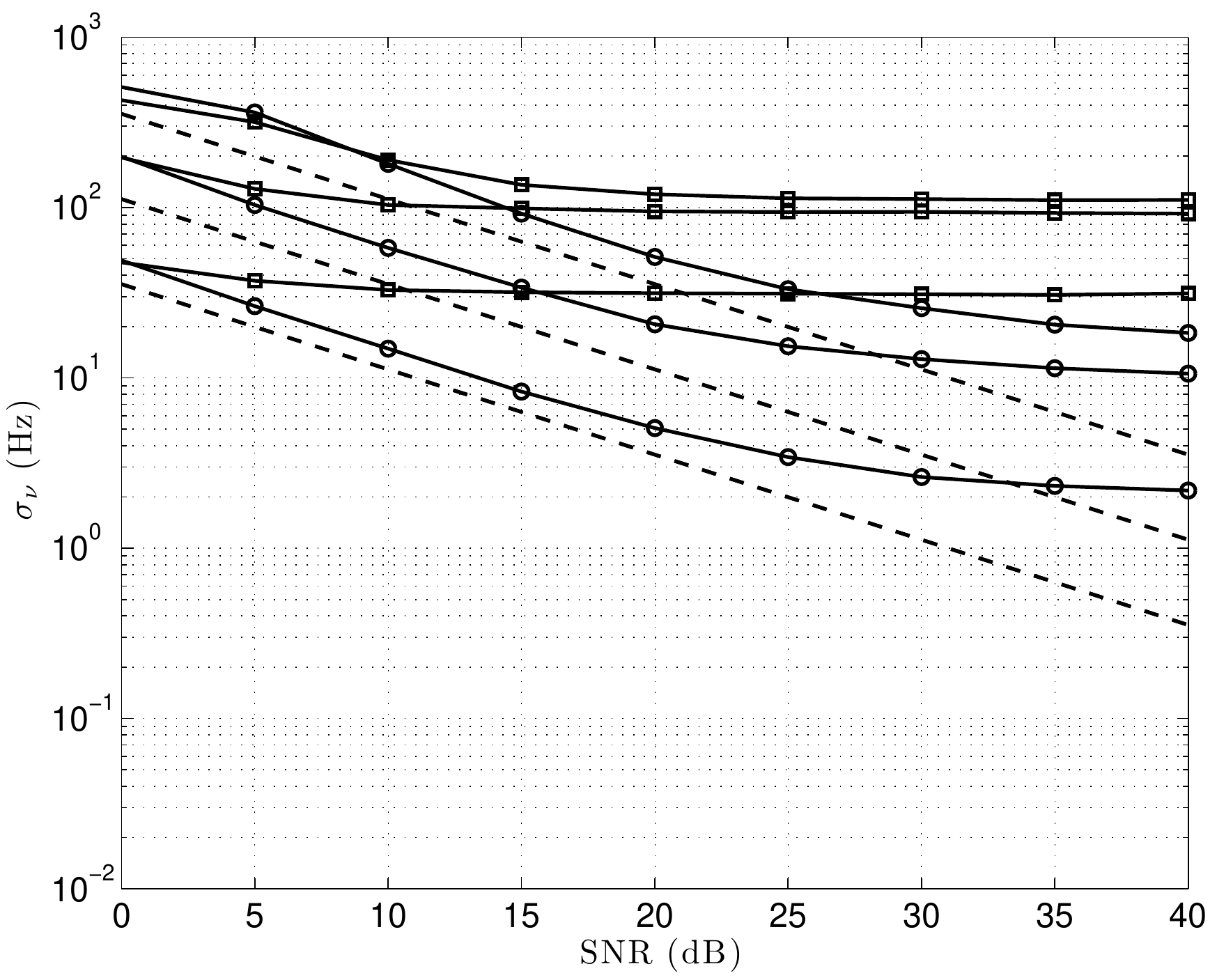} \label{fig:nu_L128_P3}}
  \subfigure[$L=256$]{\includegraphics[width=0.45\columnwidth]{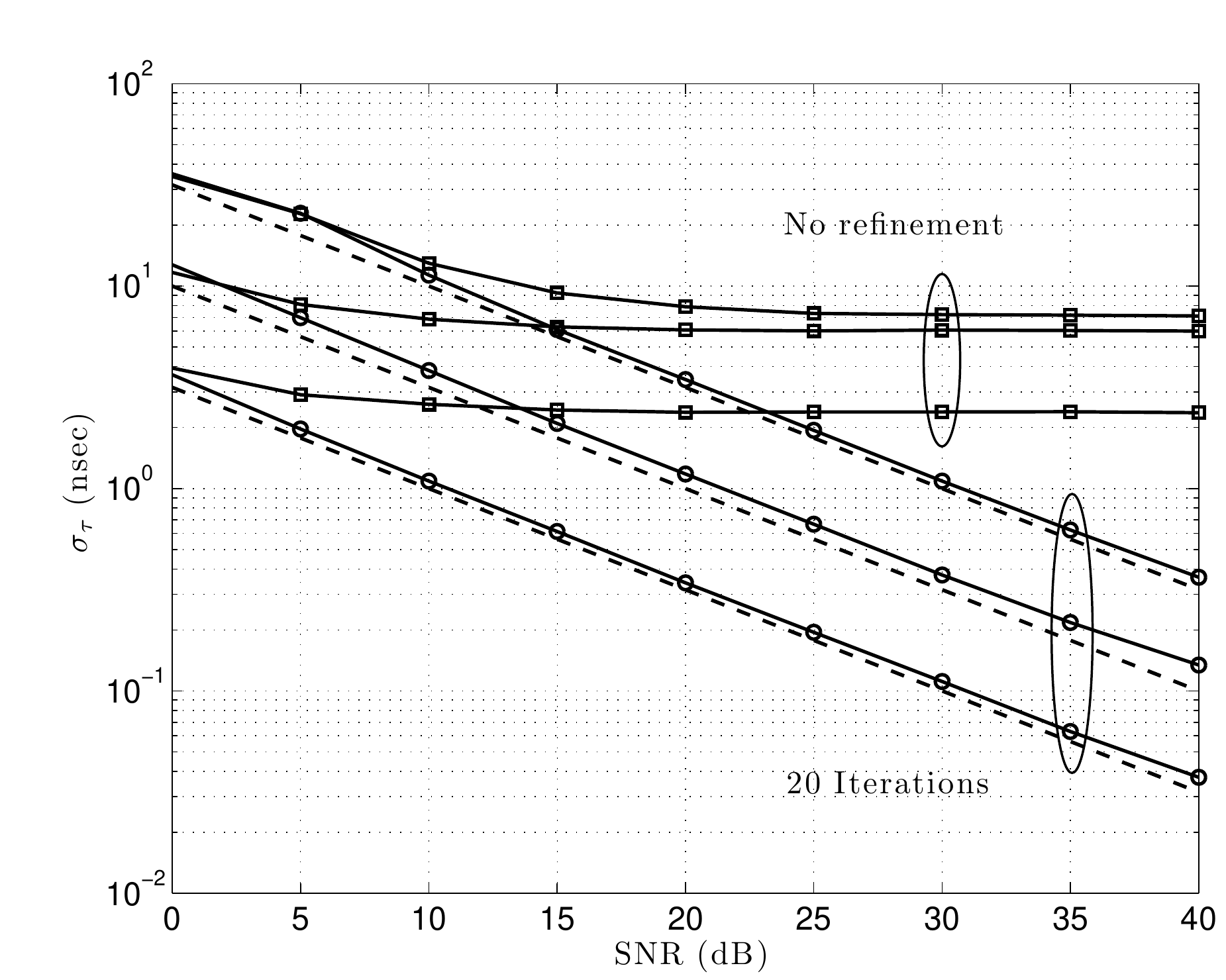}   \label{fig:tau_L256_P3}}
  \subfigure[$L=256$]{\includegraphics[width=0.45\columnwidth]{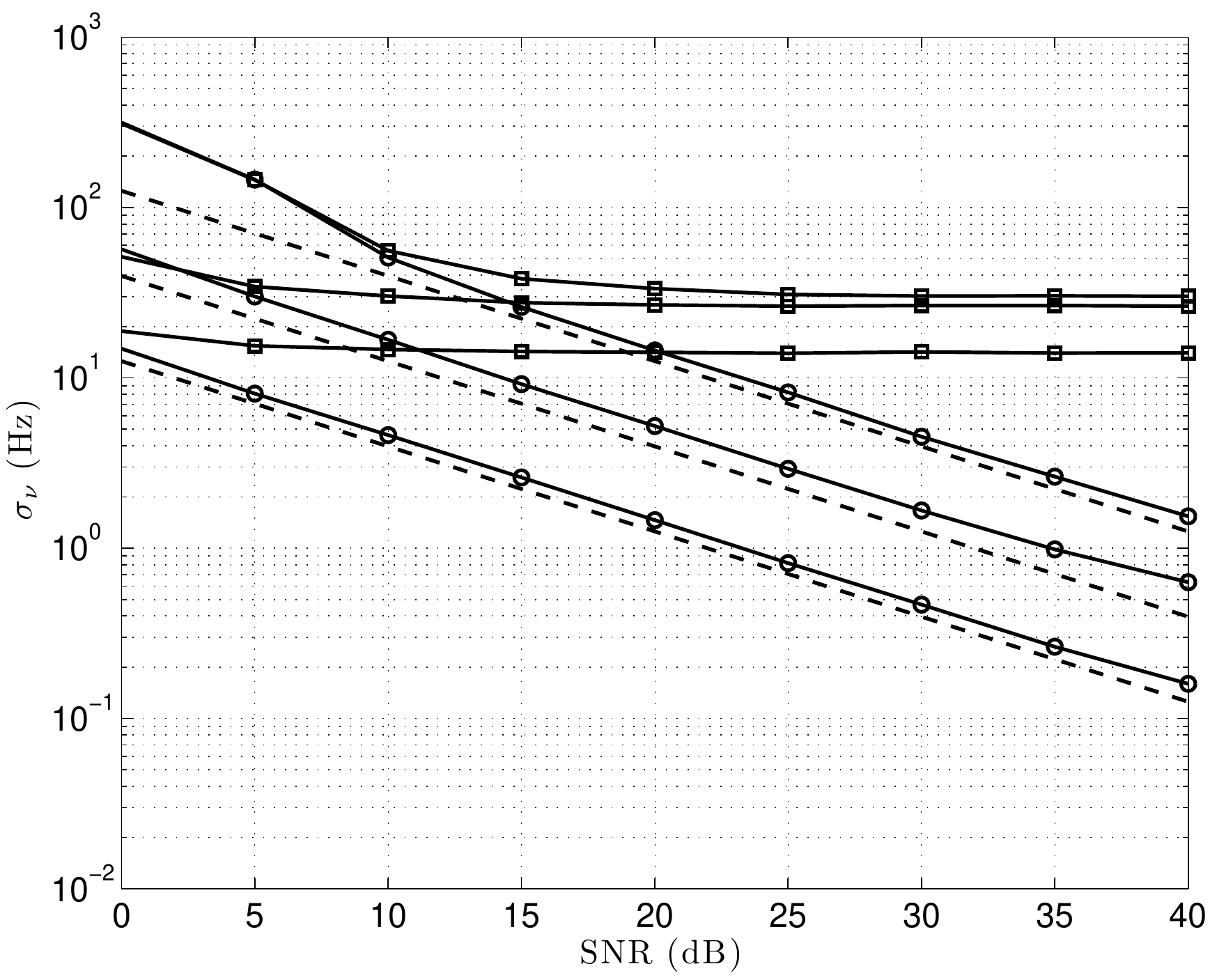}   \label{fig:nu_L256_P3}}
  \subfigure[$L=512$]{\includegraphics[width=0.45\columnwidth]{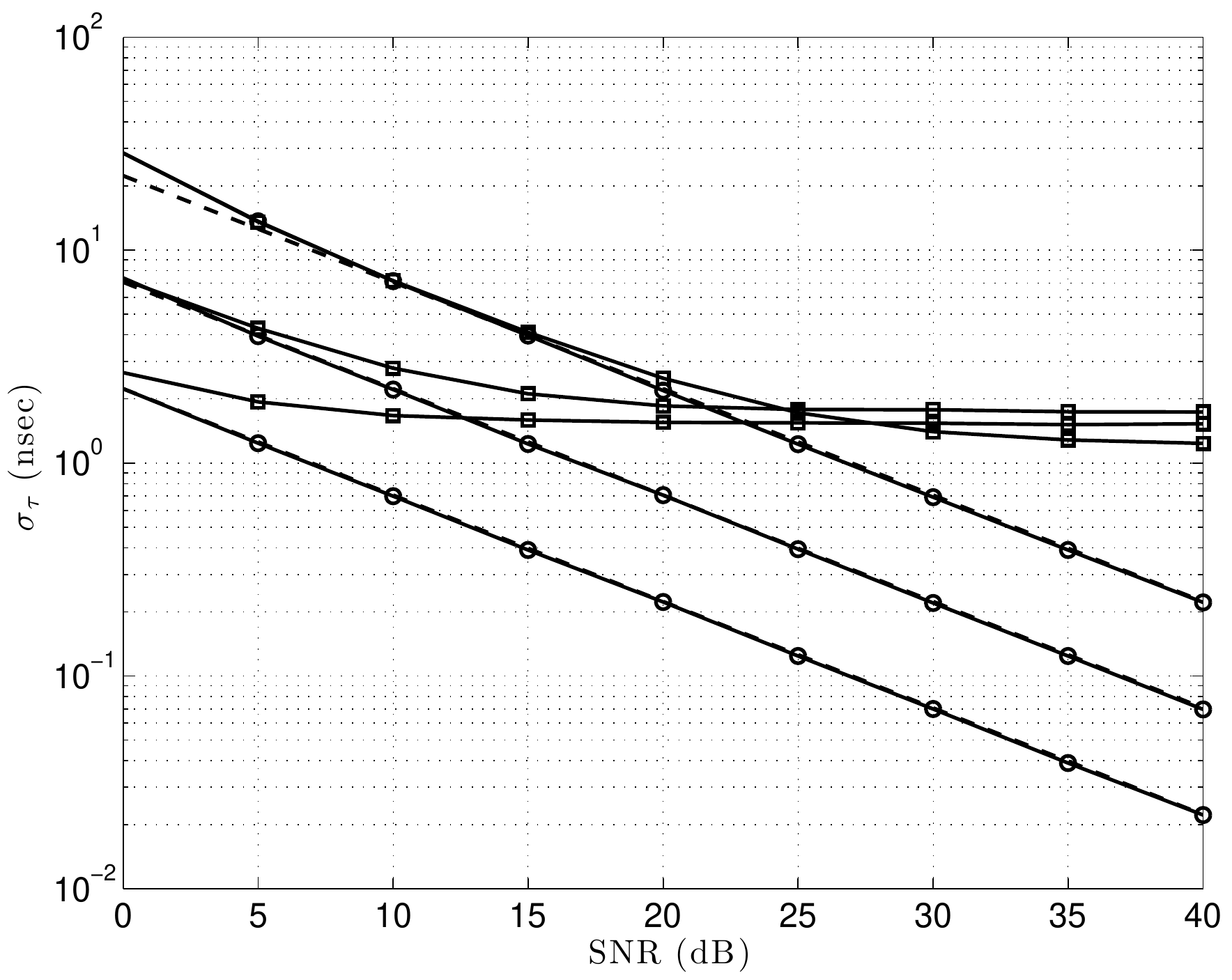}   \label{fig:tau_512_P3}}
\subfigure[$L=512$]{\includegraphics[width=0.45\columnwidth]{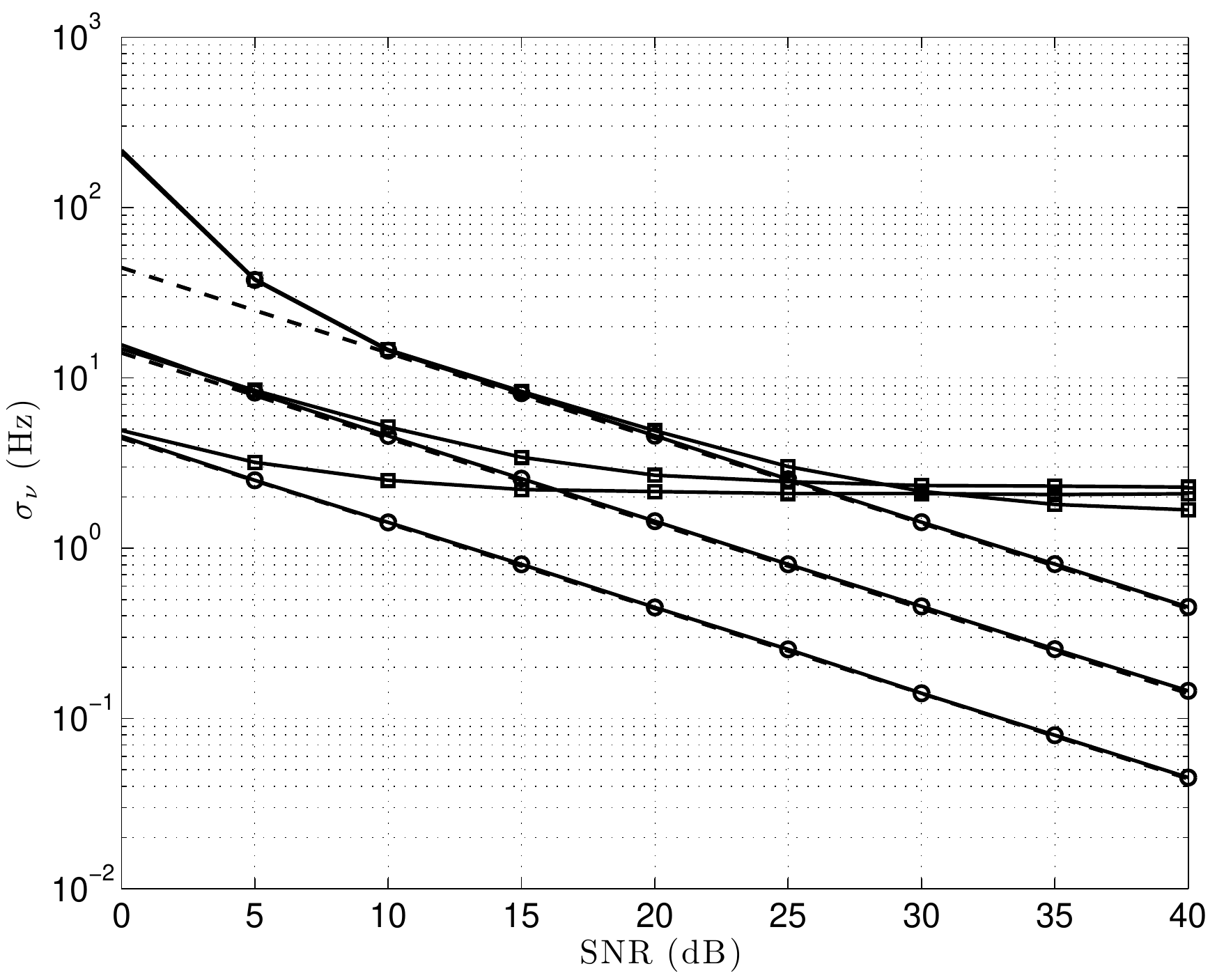}   \label{fig:nu_512_P3}}
  \caption{Three tap estimation root-mean squared (RMS) error. Dashed lines show the CRLB~\eqref{eq:crlb_single_tap} (single-tap estimation). Solid lines with squares show RMS error performance of Algorithm~\ref{alg:init_sc} only. Solid lines with circles show the RMS error performance after refinement Algorithm~\ref{alg:refine} with $20$ iterations.}
\label{fig:three_tap_rms}
\end{figure}

Fig.~\ref{fig:three_tap_rms} shows the RMS estimation error results
(recalling that this is restricted to instances where missed detection
does not occur). The square marked curves show the RMS error when no
refinement is performed, i.e. only Algorithm~\ref{alg:init_sc} is
employed. In this case a floor in the RMS error performance is
observed (caused by undetected multipath components in the successive
cancellation process). When refinement is employed, as shown by the
circle marked curves, the error floor is significantly
reduced. Moreover, as $L$ increases the floor does not occur until
very high SNRs and the RMS error performance is primarily dominated by
the CRLB~\eqref{eq:crlb_single_tap}, which is shown by the dashed
curves. Thus with sufficiently long packet length,
Algorthms~\ref{alg:init_sc} and \ref{alg:refine} deliver single-tap
performance, i.e. are able to accurately cancel the contributions of
``interfering'' taps.

It is interesting to translate the estimator performance into
range/velocity resolution. Considering $L=512$ and SNR $20$ dB, the
3-standard-deviation values for the $-20$ dB tap are $9$ ns and $15$
Hz. This corresponds to range resolution of $2.7$ m and relative
velocity resolution (at 5.9GHz) of $0.77$ m/s ($2.7$ km/h). This
clearly demonstrates the capability to accurately resolve quite
challenging multipath channels.

%
%
\section{Conclusion} \label{sec:conclusion}

In this paper we examined amplitude, delay and Doppler estimation of
the multipath channel taps from OFDM signal transmission in a doubly
selective mobile environment. Under certain practical system design
and mobile channel assumptions, we showed that the frequency domain
channel coefficients for an entire OFDM packet can be written as the
superimposition of 2-D complex sinusoids. The angular frequency of
each sinusoid is proportional to the delay and Doppler of a particular
multipath tap.

ML estimation of the delay/Doppler requires non-linear least squares
minimisation, which is computationally infeasible for practical
implementation. We therefore proposed a low complexity suboptimal
estimation method, based on successive cancellation, whereby multipath
components are removed once they are detected. The complexity
reduction results from a simplification of the channel model, where
time variations manifest only as Doppler frequency offsets for each
tap. For a single tap channel, this method is maximum likelihood. The
performance of this successive cancellation approach can be degraded by
interference from taps that are yet to be detected in future
iterations. To remedy this, we proposed a refinement algorithm based
on parallel cancellation, i.e. all estimated multipath components are
subtracted except the component of interest, which is subsequently
re-estimated.

The performance of our estimator was shown to be dominated by two
effects: separation of the multipath taps in the delay/Doppler plane;
and noise. When two or more taps are close together in the 2-D
delay/Doppler space, the estimator may detect these as single tap,
resulting in missed detections and significantly degrading the RMS
error of other detected taps.  When the multipath taps are
sufficiently separated in delay/Doppler the estimator performance is
dominated by noise and hence the RMS error of the refined estimates
are very close the CRLB of a single 2-D sinusoid in additive white
Guassian noise. We believe the missed detections are caused by the
transmit ambiguity function: broadness of the main lobe affects the
delay/Doppler resolution; and sidelobes of components that have not
been sufficiently subtracted can mask weaker taps. However, a detailed
analytic investigation of these effects is beyond the scope of this
paper and the subject of future work.

Note that although our results assume delay spreads less than the
cyclic prefix, our proposed estimator still works well without this
restriction. Multipath taps with delay exceeding the cyclic prefix
will introduce inter-symbol interference. The estimator views this
interference as extra noise on the received symbols. Thus, as long as
AWGN dominates, this extra interference will have negligible effect on
performance.

\appendices
\section{Derivation of Receiver Matched Filter Output} \label{app:rx_matched_filt}
For clarity we repeat the transmitted signal,
\begin{equation}
x(t) = \sum_{l'} x_{l'}(t) = \frac{1}{\sqrt{KL}} \sum_{l',k'} X_{l',k'} w(t - (l'-1)T_d) e^{j 2 \pi (k' - 1 - \lfloor K/2 \rfloor)(t - T_{\rm cp})/T}. \label{eq:tx_sig}
\end{equation}
Application of the channel response~\eqref{eq:chan_response} to the
transmitted signal~\eqref{eq:tx_sig} yields
\begin{align}
y(t) &= x(t) * h(t,\tau) + z(t) = \int_{-\infty}^{\infty} x(t - \tau) h(t,\tau) \, d \tau \notag \\
&=  \frac{1}{\sqrt{KL}} \sum_{l',k',p}a_p X_{l',k'} \int_{-\infty}^{\infty} g(\tau-\tau_p) w(t - \tau - (l'-1)T_d) e^{-2 \pi \nu_p t} e^{j 2 \pi (k'-1 -\lfloor K/2 \rfloor)(t - \tau - T_{\rm cp})/T} \, d\tau + z(t) \notag \\
&=  \frac{1}{\sqrt{KL}} \sum_{l',k',p}a_p X_{l',k'}e^{-j 2 \pi (k'-1-\lfloor K/2 \rfloor)T_{\rm cp}/T}  e^{-j 2 \pi \left( \nu_p - \frac{k'-1}{T} + \frac{K}{2T}\right)t}  s_{l',k'}(t,\tau_p) + z(t), \label{eq:rx_sig_generic}
\end{align}
where the integral
\begin{equation}
 s_{l',k'}(t,\tau_p) =  \int_{-\infty}^{\infty} g(\tau-\tau_p) e^{-j 2 \pi (k' - 1 -\lfloor K/2 \rfloor)\tau/T} w(t - \tau - (l'-1)T_d) \, d\tau, \label{eq:phi}
\end{equation}
is simply the convolution of a time/frequency translated filter
response $g(t)$ and time shifted window function $w(t)$. Using the
appropriate properties of Fourier transforms, the Fourier transform
of~\eqref{eq:phi} can be written as
\begin{equation}
 S_{l',k'}(f,\tau_p)  = e^{-j 2 \pi \tau_p \left[ \frac{k'-1}{T} - \frac{K}{2T} + f\right]} e^{-j 2 \pi f (l'-1) T_d} G \left(f+\frac{k'-1}{T} - \frac{K}{2T}\right) W (f),
\end{equation}
where $G(f)$ and $W(f)$ denote the Fourier transforms of $g(t)$ and
$w(t)$ respectively. In practical OFDM systems the passband bandwidth
of $G(f)$ is typically greater than $K/T$ and the bandwidth of $W(f)$
is typically less than $1/T$, e.g. for the simple case $\tilde{w}(t) =
\frac{1}{\sqrt{T_d}}$, then $W(f) = \sqrt{T_d} e^{-j \pi f T_d} \sinc
( \pi f T_d)$. Moreover, in many OFDM standards the outer subcarriers
are null subcarriers. Hence assuming negligible passband ripple then
$G \left(f+\frac{k'-1}{T} - \frac{K}{2T}\right) \approx 1$ for
$k'=1,\ldots,K$ and $|f| < 1/(2T)$. Therefore,
\begin{equation}
 S_{l',k'}(f,\tau_p)  \approx e^{-j 2 \pi \tau_p \left[ \frac{k'-1}{T} - \frac{K}{2T} + f\right]} e^{-j 2 \pi f (l'-1) T_d} W (f).
\end{equation}
Thus, taking the inverse Fourier transform yields
\begin{align}
  s_{l',k'}(t,\tau_p) & \approx  e^{-j 2 \pi \tau_p \left[ \frac{k'-1}{T} - \frac{K}{2T}\right]} \int_{-\infty}^{\infty} W (f)  e^{j 2 \pi f \left[ t - \tau_p - (l'-1) T_d\right]} \, df  \notag \\
&=  e^{-j 2 \pi \tau_p \left[ \frac{k'-1}{T} - \frac{K}{2T}\right]}  w(t - \tau_p - (l'-1) T_d). \label{eq:phi_approx}
\end{align}
Substituting~\eqref{eq:phi_approx} into~\eqref{eq:rx_sig_generic} gives,
\begin{align}
y(t) &=  \frac{1}{\sqrt{KL}} \sum_{l',k',p}a_p X_{l',k'}e^{-j 2 \pi (k'-1-\lfloor K/2 \rfloor)T_{\rm cp}/T}  e^{-j 2 \pi \left( \nu_p - \frac{k'-1}{T} + \frac{K}{2T}\right)t} \notag \\
& \;\;\;\;\; \times   e^{-j 2 \pi \tau_p \left(\frac{k'-1}{T} - \frac{K}{2T}\right)}  w(t - \tau_p - (l'-1) T_d)  + z(t), \label{eq:rx_sig}
\end{align}

The receiver now performs the matched filter to the transmitted sinusoids
(less the cyclic prefix), i.e.
\begin{align}
Y_{l,k} &= \frac{1}{\sqrt{KL}}\int_{T_{\rm cp} + (l-1) T_d}^{lT_d} y(t) w^*(t - lT_d) e^{-j 2 \pi (k-1-\lfloor K/2 \rfloor) (t-T_{\rm cp})/T} \, dt \notag \\
&=  \frac{1}{KL} \sum_{l',k',p} a_p X_{l',k'}  e^{-j 2 \pi \tau_p \left(\frac{k'-1}{T} - \frac{K}{2T}\right)} e^{-j 2 \pi (k'-k)T_{\rm cp}/T}   \notag \\
& \;\;\;\;\; \times \int_{T_{\rm cp} + (l-1)T_d}^{lT_d}   w(t - \tau_p - (l'-1) T_d) w^*( t - (l-1)T_d)  e^{-j 2 \pi \left( \nu_p + \frac{k-k'}{T} \right) t} \, dt + Z_{l,k} \notag \\
&=  \frac{1}{KL} \sum_{l',k',p} a_p X_{l',k'}  e^{-j 2 \pi \tau_p \left(\frac{k'-1}{T} - \frac{K}{2T}\right)} e^{-j 2 \pi (k'-k)T_{\rm cp}/T} e^{-j 2 \pi \left( \nu_p + \frac{k-k'}{T} \right) (l-1)T_d}  \notag \\
& \;\;\;\;\; \times \hat{A}_w\left( \tau_p + (l'-l) T_d,  \nu_p + \frac{k-k'}{T} \right) + Z_{l,k}, \label{eq:ofdm_rx_mf}
\end{align}
where
\begin{equation}
\hat{A}_w(\tau,\nu) =  \int_{T_{\rm cp}}^{T_d} w(t - \tau)  w^*(t) e^{-j 2 \pi \nu t}  \, d t,
\end{equation}
which resembles the ambiguity function of $w(t)$.\footnote{The
  function $\hat{A}_w(\tau,\nu)$ is not quite the ambiguity function
  of $w(t)$ because of the limits of integration.}  In practical OFDM
systems, usually $\max_{p} \tau_p < T_{\rm cp}$ and $\max_p \nu_p \ll
1/T$ and the windowing function is usually designed such that \\
$\hat{A}_w\left(\tau_p + (l'-l)T_d, \nu_p + \frac{k-k'}{T} \right)
\approx 0$ for $k \neq k'$ or $l \neq l'$. 
Hence we may write,
\begin{align}
Y_{l,k} &=  \frac{1}{KL} \sum_{p} a_p X_{l,k}  e^{-j 2 \pi \tau_p \left(\frac{k-1}{T} - \frac{K}{2T}\right)} e^{-j 2 \pi \nu_p T_d  (l-1)} \hat{A}_w ( \tau_p,\nu_p ) + Z_{l,k} \notag \\
&=  \frac{1}{KL} \sum_{p} \tilde{a}_p X_{l,k}  e^{-j 2 \pi (k-1) \tau_p/T } e^{-j 2 \pi (l-1)\nu_p T_d }  + Z_{l,k},
\end{align}
where $\tilde{a}_p = e^{-j \pi K \tau_p/T}
\hat{A}_w(\tau_p,\nu_p)$. With some slight abuse of notation, for the
remainder of the paper for brevity of notation (and without loss of
generality) we will replace $\tilde{a}_p$ with $a_p$. Defining $H_{l,k}$ according to \eqref{eq:channel_coeffs}, we obtain~\eqref{eq:chan_model}. 



\end{document}